\documentclass{article}
\usepackage{graphicx}
\usepackage{amsmath}
\usepackage[cm]{fullpage}
\usepackage{hyperref}
\usepackage{multicol}
\usepackage[utf8]{inputenc}
\usepackage{array}
\usepackage{enumerate}
\usepackage{multicol}
\usepackage{microtype} % reduce hypenation
\usepackage{tikz}

\usetikzlibrary{decorations.pathmorphing,patterns,shapes,arrows,positioning,decorations.pathreplacing,calc,mindmap,intersections}

\newcolumntype{L}[1]{>{\raggedright\let\newline\\\arraybackslash\hspace{0pt}}m{#1}}

\title{Forecasting Bus and Station Occupation in a Public Transport Ridership Model lacking Route or Alighting Location, the case of TransMilenio in Bogotá}

\author{Gabriel Villalobos\footnote{Corresponding Author. \emph{Laboratorio de Innovación en Administración Pública, Escuela Superior de Administración Pública ESAP}, Address: Calle 44 \# 53 - 37, Bogotá, Colombia ZIP 111321. \url{gabriel.villalobos@esap.edu.co}. }   Juan D. Garcia-Arteaga\footnote{\emph{Facultad de Medicina, Universidad Nacional de Colombia} Calle 45 Carrera 30, Bogota, Colombia, ZIP 111321, \url{judgarciaar@unal.edu.co} }  Arturo Argüelles\footnote{\emph{Departamento de Física, Facultad de Ciencias Naturales y Exactas, Universidad del Valle}, Calle 13 \#100-00, Cali, Colombia. \url{arturo.arguelles@correounivalle.edu.co}} 
}

\tikzstyle{decision} = [diamond, draw, fill=yellow!20, 
    text width=4.5em, text badly centered, node distance=2.5cm, inner sep=0pt]
\tikzstyle{block} = [rectangle, draw, fill=blue!20, node distance=2.5cm, 
    text width=5em, text centered, rounded corners, minimum height=4em]
\tikzstyle{orangeblock} = [rectangle, draw, fill=orange!50, node distance=2.5cm,    text width=5em, text centered, rounded corners, minimum height=4em]
\tikzstyle{charblock} = [rectangle, draw, fill=teal!20, node distance=2.5cm, 
    text width=5em, text centered, minimum height=4em]
\tikzstyle{funcblock} = [rectangle, draw, fill=red!20, node distance=2.5cm, 
    text width=5em, text centered, rounded corners, minimum height=4em]
\tikzstyle{line} = [draw, -latex']
\tikzstyle{cloud} = [draw, ellipse,fill=red!20, node distance=2.5cm,
  minimum height=2em]
 \tikzstyle{io} = [trapezium, draw, trapezium right angle=110, rounded corners, fill=red!20, node distance=1.9cm, minimum height=1.5em]    

\begin{document}
\maketitle

\begin{abstract}
  {We present a detailed ridership model for a public transport system
    that can continuously forecast user waiting times as well as bus
    and station occupancy based on automatic fare collection (AFC)
    data. These models usually have one of two restrictions. For
    subway/metro systems the entry and exit stations can be known, but
    the specific route within the system is unknown; for bus systems
    usually the exit station is unknown, but the route is known.
    
    Bus Rapid Transit systems share some elements of bus and
    subway/metro systems, which makes them flexible options for public
    transport. Moreover, this also implies that they can not be modeled
    as either of them.  Take the case of TransMilenio, the Bus Rapid
    Transit (BRT) system that is the backbone of public transport in
    Bogotá. Users pay a flat fare that does not depend on the length
    of the trip, and the ticket is validated on at the entry station,
    not on bus. Therefore, TransMilenio data has both restrictions; it
    lacks the information on the alight location and the
    particular route taken by users.  We propose a model that allows
    to forecast bus and station occupation in such transit systems;
    and show an example of implementation of the model over the
    \emph{troncal} component of TransMilenio.

  We compare the average bus occupation of two particular days: 17 of
  May 2020, during the sanitary crisis caused by Covid-19; and on
  December 1, 2020, after most of the mobility restrictions had been
  lifted in Bogotá. In the former, very rarely the average bus
  occupation distribution exceeds 0.5; while in the latter most buses
  exceed this threshold. The model can be used to improve the quality
  of public transport, as well as to reduce costs of operation.}
\end{abstract}

% \begin{keyword} Ridership
%   Model \sep TransMilenio \sep Bus rapid transit \sep bus occupancy \sep Network Model
%   \end{keyword}

%\begin{document}

%\maketitle \abstract

%\begin{multicols}{2}
           
\section{Introduction}

%\subsection{Literature review}
Transit systems are able to store usage information by means of
automatic fare collection (AFC). Users validate their tickets either
before entering the station or from within the system itself.  In
principle the granularity of the AFC data could be aggregated to have
a detailed picture of the usage of the system; with the goal of
estimating the temporal and spacial passenger flow distribution spread
over the system.

If both entry and exit locations are recorded in the AFC data the
origin destination (OD) matrices can be constructed. These matrices
can be classified as time dependent or equilibrium, and in a
simulation model could be used to generate the passengers as discussed
by Yao et al. \cite{yao2017simulation}. A possible way to generate the
OD matrix is to utilize the GPS information of the buses and the AFC
information, as done by Munizaga and Palma \cite{MUNIZAGA20129}.

According to Si et al. \cite{doi:10.1061/JTEPBS.0000444}, models to
solve the urban transit assignment problem can be divided in two
categories. Firstly the frequency-based models, in which the average
section flow through sections of the urban network are calculated with
long-term planning in mind. Secondly the schedule-based models, in
which the activity of each passenger is considered; which renders
information for specific planning.

There are two general sources of uncertainty on the granularity of the
AFC data. Firstly, the AFC may not record which particular route the
user took within the system; as is the case in most subway or metro
systems \cite{https://doi.org/10.1002/atr.1309}.
The second uncertainty in the AFC information is known as the
destination inference problem; which is the situation in which the
route within the system is known but the initial or final stations are
unknown. It is usually the case for bus systems in which the
validation is taken inside the bus.

The mass transit systems based on buses (also known as bus rapid
transit or BRT) share some characteristics with both subway/metro and
bus systems. In terms of the information available for modeling, they
may both lack the information on the selected route and face
the destination inference problem. Moreover, equilibrium models may not
be suitable, due to the uncertainty due to traffic even in dedicated
streets.

In this paper we present a schedule-based simulation model for a BRT
system. The main contributions of our work are:

\begin{itemize}
  \item It acknowledges for two kinds of uncertainty, both on the
    route choice and the destination inference. Therefore, can
    be used for BRT systems who share characteristics of both metro
    and bus systems.
\item It implements the network loading model based on AFC data,
  providing the temporal and spatial flow distribution spread over the
  system.
\end{itemize}

Our model can tackle both the route uncertainty and the
destination inference problems. The goal with the model is to estimate
temporal and spacial passenger flow distribution, taking into account
the capacity of the buses. As an application we implement it for the
case of TransMilenio, the BRT that constitutes the backbone of the public transport system of the city of Bogotá.

In the remainder of the introduction we present a literature review of
the transit assignment models. We also go through the specific
literature related to TransMilenio. In the next section we present a
description of the model and the source of data that is used to make
the simulation. After that we discuss the results of the model and
present further work.

\subsection{Literature Review of Transit Assignment Models and Route Choice}

In the schedule-based model by Tong and Wong,
\cite{https://doi.org/10.1002/atr.5670330307}, the transit system is
represented by a graph with nodes links and lines. Nodes represent the
source or sink of passengers, links can be either transit or walk
links among the nodes, and lines are the paths taken by transit
vehicles. Given the network a minimum path algorithm that takes into
account walking time, in-vehicle time and line change penalty is
calculated. The optimal path is obtained with Dijkstra's
algorithm \cite{dijksta1959note}. This model assumes fixed schedules, and does not have a
capacity restraint provision in the buses, queuing at the stations or
delay due to congestion in the departing of the buses. Moreover, assumes
that passengers choose the minimum weighted time path. Tian and Huang
extended the model including capacity constrains
\cite{tian2007commuting} for a monocentric city, and forcing the
equilibrium condition in which commuters departing form the same
station have identical generalized travel cost. Si et al. present an
equilibrium model with similar conditions, and implement it for
Beijing \cite{doi:10.1061/JTEPBS.0000444}.

Yao et al. make the distinction between mathematical models and
simulation models, the former being oriented at equilibrium
considerations and the latter towards more dynamic aspects of real
flows of passengers \cite{yao2017simulation}. Simulation models also
treat individual passengers as agents within the simulation. They
place these models within the category of Dynamic Traffic Management
(DMS) and Dynamic Traffic Assignment (DTA). In their view a DTA has
two components, the route selection model and the network loading
model. The former describes how passengers choose the route, the
latter represents the evolution of passenger flows on the network. They
also introduce the concept of generating the passengers within the
simulation out of the AFC data. 

Regarding the problem of choosing a particular path within metro
systems, Zhang et al. presented a model that combines information
obtained from questionnaire surveys -which they call Revealed
Preference- and AFC data \cite{ZHANG201876}. Wu and Dong study the
performance of metro networks \cite{WU2019787}. Their model fixes a
routing strategy that optimizes by the number of transfers, and they
study the robustness of the network by means of a proposed
centrality: the node occupying probability. Cheng et al. propose a
method to estimate the route choice by means of analyzing the
statistical properties of the distribution of travel times and OD
matrices obtained from AFC data
\cite{doi:10.1177/0142331218823855}. Gordon et al. work on the
estimation of OD matrices in the cases for which there is
heterogeneity in the methods of payment; for instance if there is a
mix of AFC, magnetic cards, and cash \cite{GORDON2018350}.

Other aspects that have been studied in the literature of public
transit systems include ridership prediction, which refers to
forecasting the usage of the public transit systems and their
reliability \cite{7954025,https://doi.org/10.1002/atr.1332}; as well
as passenger flow within the system,
\cite{doi:10.1080/21680566.2018.1434020}; the distribution of time
between passenger arrival, \cite{https://doi.org/10.1111/mice.12265};
and usage of OD matrices to match between demand and supply in
transport planning using data,
\cite{9091900,BEHARA2021103370}. Moreover, passenger flow simulations
follow individuals through the transport system to understand the
passenger demand and its relation with the schedules of the systems
\cite{cai2021evaluation,chen2019mas,chen2019mas}. An example of the
solution of the destination inference problem in a bus system with
route information is given by Trepanier et al. \cite{doi:10.1080/15472450601122256}.

\subsection{TransMilenio}

BRT systems such as TransMilenio in Bogotá can be
implemented as an alternative to metro/subway systems. They share some
characteristics, services run through lines that stop at stations,
some of which serve as transfer between lines. Users enter the system
at a given station, using their AFC card. The main differences are in
capacity of the bus compared to the train, and the absence of complete
subway/metro lines. TransMilenio has dedicated lines for their BRT
buses, but they do interact with other forms of vehicle traffic at
intersections. It also uses a flat rate for usage of the system,
therefore the AFC card is not used to exit the system; which for the
modeler implies that we need to solve the route choice and the
destination inference problem.

The availability of automatic fare collection from TransMilenio has
helped studies covering fare discrimination
\cite{guzman2018fare,guzman2020short,GUZMAN2021335,GUZMAN2021103071},
out-of-home participation of households \cite{COMBS201711}, as well as
social fragmentation and relationship to density, land use, and land
value
\cite{doi:10.1177/0042098015588739,BOCAREJO201378,RODRIGUEZ20164}. The
congestion in the stations and comfort within the system has been
studied by surveys \cite{guerra2013congestion,pinzon2019analisis}. To
the best of our knowledge there are no models for the congestion of
individual lines or buses in TransMilenio.

\section{Materials and Methods}

We implement a schedule-based simulation model with a multiple node class
network model for the buses, a destination inference based on symmetry
of trips, and generate the passengers of any given day with AFC data
that is publicly available. 

\subsection{Network Models in Mass Transit}

TransMilenio is a complex system in which multiple agents and entities
interact. To draw useful conclusions it is necessary to model it by
reducing it to its essential information. Networks provide an
excellent theoretical framework for these types of systems.

A network is a mathematical object $G$ containing two collection of objects: Vertices $V$ and edges $E$.  Each edge connecting two vertices 
marks a relationship between them. In the area of transportation and urban planning vertices typically represent physical locations and edges 
represent connections between them \cite{newman2018networks}, e.g. airports and airline routes or cities and highways. Networks are interchangeably called graphs in mathematical contexts and especially in graph theory, an active and rich field of study.

The traditional network model used in transportation problems is the Origin Destination Matrix (OD). To construct it cities are divided into predetermined zones and the outgoing and incoming trips are estimated either by direct measurements such as traffic sensors or by indirect methods such as polls. A matrix in which each row corresponds
to the geographical origin of a trip and each column corresponds to
the destination of that trip is then constructed. The OD may be then interpreted as the adjacency matrix of a directed graph with vertices corresponding to the predetermined
zones. In this case edges have a direction (origin to destination) and a weight (number of trips). 

A more nuanced network model may include each of the legs of a trip and not simply the ``main'' destination, usually a home-work-home round trip. The data collected by
TransMilenio may allow us to do such a model in which each node
corresponds to a station and the edges correspond to the number of
travelers assumed to travel between the stations in a given direction\cite{garcia2023network}.

A third approach is to include aditional information of the operation of the system such as bus schedules and routes. This allows one to make predictions about occupancy and gives a more realistic view about the passengers' use of the system.  In the next section we will explain the construction of this model from available TM data.

\subsection{Transmilenio hybrid node model}

We have modeled the TM system by a heterogenous graph, \textit{i.e.} a graph with different node or edge types. Our model has directed edges and contains two types of nodes corresponding to stations ($S$) and buses ($B$): 
%The main characteristic of bipartite graphs is that each node has one of two labels and that edges are only allowedbetween nodes with different labels. 
%For TransMilenio we have built a
% graph with two node types: 

 $$G=( (V_s,V_b),E)$$

\begin{figure}[htbp]
  \includegraphics[width= \columnwidth]{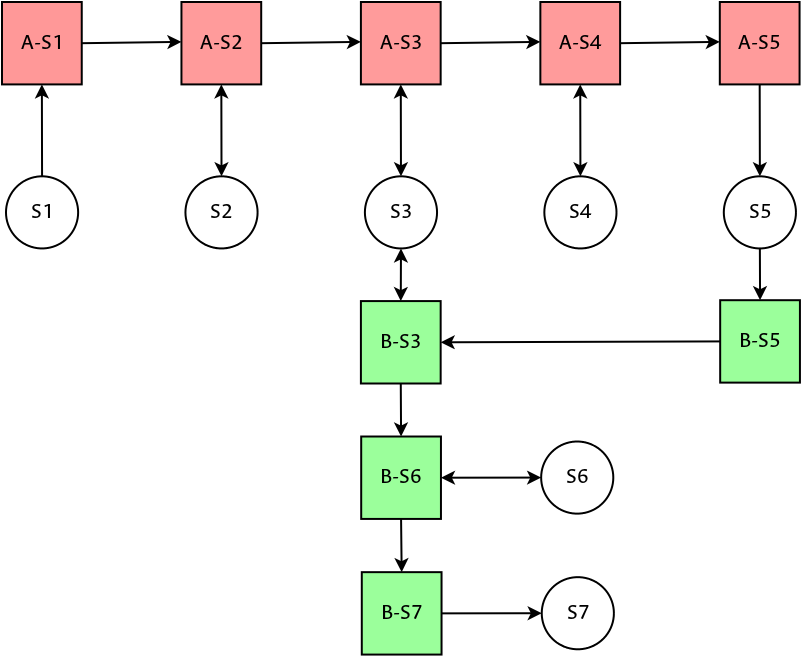}
  \caption{A simplified model of TransMilenio as a network with two node types: Square nodes correspond to bus lines and circular nodes correspond to stations.
A traveler may only board a bus from a station. Each color corresponds to a route.}
  \label{fig:enrutamiento}
\end{figure}

TM stations are represented by vertices of type $v_s$ and bus nodes of type $v_b$. A subset of weakly connected $v_b$ nodes is called a route. Each of the nodes of a route represents a planned stop at a specific station and is connected to that station. 
%Each route node is connected as well to the previous and the next stop except for the initial and final stops 
%There are no edges between $V_s$ nodes and a route stopping at a given station is represented by an edge between one of its $v_b$ nodes and the $v_s$ node corresponding to that station. 
Except for the first and last station of the route, all edges between station and bus nodes are bidirectional as passengers may either mount or alight at them.  At the first station the edge is directed from the $v_s$ station node to the $v_b$ bus node (as passengers may only mount) and at the last station from the $v_b$ bus node to the $v_s$ station node as passengers may only alight. A $v_b$ node is connected to the next bus node in the route representing the bus movement. Figure \ref{fig:enrutamiento} shows a toy example with seven $s$ nodes (stations) and two routes (A and B). 

Using the heterogenous graph model, together with the scheduling information of the routes and the boarding time of passengers, it is possible to assign a route to each passenger
that minimizes their traveling time. We have adapted the Dijikstra shortest path algorithm to minimize the expected time of arrival of the traveler. This allows us to calculate
the occupation of a bus from a particular route at any given time. 
\subsection{Dijkstra's Algorithm}

Dijkstra's Algorithm (DA) finds the shortest path, i.e. the one with the lowest edge sum, between two nodes in a weighted graph.  The algoritm guarantees that the minimal cost will be found for graphs with non-negative weights.

In its basic form, DA marks all nodes as unvisited and assigns a tentative distance of infinity to all except the starting node which is assigned 0. A priority queue in which all unvisited nodes are placed in ascending order according to their tentative distances is intialized. The algorithm then visits the first node of the queue and assigns to its neighbors the minimum between their current tentative distance and the tentative distance of the visited node and the edge between them. When all the neighbouring nodes have been updated, the algorithm marks the node as visited and reorders the queue. The algorithm ends either when the desired destination node is marked as visited, in which case the tentative distance of the destination node is the minimum distance from the origin node, or when the priority queue is empty, in which case no path exists between the origin and the destination node. 

DA may be used in both directed and undirected static networks as long as the edges are strictly positive. Unfortunately, if we try to model the passenger's decision as one that minimizes the total time spent in the system the graph's edge weights become dynamic depending on the hour.  Additionally, TM has multiple routes with varying numbers of stops ranging from basic routes (which stop at all stations) to express routes (which make very few stops). 

To solve this, edge weights are assigned as follows.  Edges from $v_b$ to $v_s$ nodes are assigned an arbitrary cost of 1 minute as alight time.  Edges between $v_b$ nodes are assigned a time cost equivalent to the time necessary to traverse the distance between them using the speeds calculated in section \ref{Sec:PassengerFlowModel}. Finally, the cost of mounting from $v_s$ to $v_b$ is calculated as the time between the estimated time of arrival in the station, either by entering the system or descending from another bus, and the next departure of the bus. To do this efficiently a table with all estimated arrival/departure times is assigned to each node and calculated using the departure schedules and the distance between stations. 

The DA calculates ideal trajectories for users which would only work under the assumption that all users have perfect information and can calculate optimal routes. A more realistic model, beyond the scope of the present work, is to assign user trajectories based on minimal number of bus transfers or minimal average waiting time.

\subsection{Passenger Flow Model}
\label{Sec:PassengerFlowModel}

The passenger flow model calculates the dynamics of the users on the
nodes of the mass transit network. Take a bipartite graph representing
the network as $G(N,A)$, where $N$ are the nodes and $A$ the
links. Nodes can be either stations, $N_{S_{i}},N_{S_{ii}},...$; or
lines $N_{L_{i}}, ...$. 

With the model introduced in the previous section we calculate the
path through the network that minimizes the travel time for a
passenger at different moments in time. The model takes into account
the schedule and capacity of the line. We use the word 'bus' in our
model to refer to a service that runs through a line at a particular
time. The model allows us to assign to each passenger the ordered set
of ordered nodes to go through. If for instance it may start at
station $i$, take line $i$, goes to station $vi$, take line $iv$, and
end at station $xiv$: $\{N_{S_{i}} \rightarrow N_{L_{ii}} \rightarrow
N_{S_{vi}} \rightarrow N_{L_{iv}} \rightarrow N_{S_{xiv}} \}$.

The system evolution follows the dynamics of an event driven
simulation discrete time (seconds) since the beginning of the
day. The main events that generate the change of the internal state of
the agents in the system is the arrival of a bus to a station.

\begin{figure}[htbp]
  \center
    \resizebox{0.8 \columnwidth}{!}{
\begin{tikzpicture}[small mindmap,concept color=teal!40,grow cyclic,level 1/.append style={level distance=3cm,sibling angle=35}]

  \node[cloud] (inicio) {Start};
  \node [io, below of= inicio](leeEstaciones){Read Static TM Data\\ \vspace{3mm}};
  \node [charblock, below of= leeEstaciones](generaRutas){Generate Buses and Routes};
  \node [io, below of= generaRutas](leeUsuarias){Read Daily Ridership Data};
  \node [charblock, below of= leeUsuarias](generaUsuarias){Generate Users \\ Set time counter to 0};
  \node [decision, below of= generaUsuarias](loopT){System is still running? };
  \node [decision, below right of= loopT](avanzaBus){next bus\\arrived to an estation?};
  \node [block, right=0.7  of avanzaBus](itinerary){Update Users in bus};
  \node [block, below  of= itinerary](itinerary2){Update Users in Station};
  \node [cloud, below left of= loopT](end){End};
  \node [io, below of= avanzaBus](printData){Write Bus and Station Data};

  \path [line] (inicio) -- (leeEstaciones);
  \path [line] (leeEstaciones) -- (generaRutas);
  \path [line] (generaRutas) -- (leeUsuarias);
  \path [line] (leeUsuarias) -- (generaUsuarias);
  \path [line] (generaUsuarias) -- (loopT);
  \path [line] (loopT) -| node [text width=2.5cm,midway,above=-0.9cm,align=center ]{Yes} (avanzaBus);
  \path [line] (avanzaBus) -- node [text width=2.5cm,midway,above=-0.9cm,align=center ]{Yes} (itinerary);
  \path [line] (itinerary) -- (itinerary2);
  \path [line] (itinerary2) -- (printData);
%  \path [line] (itinerary) |- (printData);
%  \path [line] (avanzaBus) -- (printData);
  \path [line] (printData) -| (loopT);
  \path [line] (loopT) -| (end);
  \path [line] (avanzaBus) -- node [text width=2.5cm,midway,above=-0.9cm,align=center ]{No} (loopT);
\end{tikzpicture}}
\caption{Flowchart of the model. The subroutine ``update users in
  bus'' is explained in Figure \ref{fig:UpdateUsuariasEnBus}, and the
  subroutine ``update users in station'' in Figure
  \ref{fig:UpdateUsersInStation}.}
\label{fig:flowmodel}
\end{figure}

The flowchart of the model is represented in Figure
\ref{fig:flowmodel}. Firstly, the recorded information of the system
(location of the stations, list of stations within the lines) is read
out from public TransMilenio records. With this we internally generate
stations, buses, and their lines. Empty buses are assigned to their
corresponding lines and their initial station arrival time is set up
according to the public TransMilenio schedule. Then we read the
ridership data, including optimal user trajectories that were obtained
by the network model.  The changes of the optimal trajectory that
result from differences in arrival times of buses due to the different
schedules at different moments of the day makes almost every optimal
individual trajectory unique.
% Lo revisé contando las trayectorias unicas con 37_reviso_tiempos.py

With this information we
schedule the arrival of each user agent to the station in which they enter the system according to the timestamp of their AFC data.
The system clock runs through one day of operation of the system. 
Bus
agents follow the trajectory in their lines within the system with a
given average speed (randomly assigned for each bus from a gamma
distribution with mean 15 km/h), until they arrive to their next
station.

The arrival of a bus to a station triggers the updating of the state
of those users within the bus that arrived to their station, explained
in Figure \ref{fig:UpdateUsuariasEnBus}. For each user in the bus we find out
whether it leaves this bus. If they finished their trip they are taken
out of the model, otherwise their status is updated.

Next we go through a subroutine that updates the status of users
within the station, explained in \ref{fig:UpdateUsersInStation}. It
runs over each user at the station. First it checks if it has arrived
to the station with the AFC data. For those, checks whether they want to take this
route; if affirmative gets queued to get in the bus, as the bus capacity and occupation affect whether it can take any more passengers. After it runs
through every user at the station, it runs through users in the queue,
to allow them to go into the bus; until it is full.

\begin{figure}[htbp]
    \center
  \resizebox{ \columnwidth}{!}{
  \begin{tikzpicture}[small mindmap,concept color=teal!40,grow cyclic,level 1/.append style={level distance=3cm,sibling angle=35}]
    \node[cloud] (inicio) {Start subroutine};
    \node [decision, below of= inicio](snuib){Select user in bus};
    \node [decision, below right=0.7cm of snuib](ulb){User leaves bus};
    \node [charblock, below=0.7cm of ulb](muoob){Move user out of Bus};
    \node [decision, below of= muoob](uet){User ended trip?};
    \node [charblock,   right = 0.7 cm of inicio](ulsei){User leaves system, erase it};
    \node [charblock,  left of= muoob](unsiu){Update next line in user};
    \node[cloud, left=0.5cm of snuib](end) {End};

    \path [line] (ulsei) --  (snuib);
    \path [line] (inicio) --  (snuib);
    \path [line] (snuib) -| node [text width=2.5cm,midway,above=-0.9cm,align=center ]{next user} (ulb);
    \path [line] (ulb) --  node [text width=2.5cm,midway,right=-0.01cm,align=center ]{Yes} (muoob);
    \path [line] (muoob) --  (uet);
    \path [line] (uet) -| node [text width=2.5cm,midway,right=-0.9cm,align=center ]{Yes} (ulsei);
    \path [line] (uet) -| node [text width=2.5cm,midway,above=-0.9cm,align=center ]{No} (unsiu);
    \path [line] (unsiu) --  (snuib);
    \path [line] (snuib) -- node [text width=2.5cm,midway,above=-0.7cm,align=center ]{No more \\users} (end);
    \path [line] (ulb) -- node [text width=2.5cm,midway,above=-0.9cm,align=center ]{No}  (snuib);

  \end{tikzpicture}}
  \caption{Subroutine: Update Users in Bus}
  \label{fig:UpdateUsuariasEnBus}
\end{figure}

\begin{figure}[htbp]
    \center
  \resizebox{ \columnwidth}{!}{
  \begin{tikzpicture}[small mindmap,concept color=teal!40,grow cyclic,level 1/.append style={level distance=3cm,sibling angle=35}]
    \node[cloud] (inicio) {Start subroutine};

%    \node [block, right of= llegoAlaSiguiente](actualizoEstacion){For users in station};
%    \node [decision, below of= actualizoEstacion](bajoUsuarias){User goes off the bus here?};

%    \node [block,  right of= actualizoEstacion](RecargaUsuaria){Change user status};
%    \node [decision, below right of= RecargaUsuaria](tomanOtro){User takes another route?};
%    \node [block, below of= tomanOtro](sebaja){User leaves Station};
    
    \node [decision, below of= inicio](actualizoUsuarias){For each user at this Station};
    \node [decision, below right of= actualizoUsuarias](siYallego){Has arrived already};
    \node [decision, below right of= siYallego](meSirve){wants to take this route?};
    \node [charblock, below of = meSirve](subalo){queue user to get on the bus};
    \node [decision, below left of = actualizoUsuarias](sobrequierensubir){For users in queue};
    \node [decision, below left of = sobrequierensubir](quepo){space on bus?};
    \node [charblock, below of = quepo](subete){get user on bus};

    \node[cloud,  below left of = quepo](end) {End subroutine};
    \path [line] (inicio) -- (actualizoUsuarias);
    \path [line] (actualizoUsuarias) -| node [text width=2.5cm,midway,above=-0.7cm,align=center ] {next User} (siYallego);
    \path [line] (siYallego) --  node [text width=2.5cm,midway,above=-0.8cm,align=center ] {No} (actualizoUsuarias);
    \path [line] (actualizoUsuarias) --  (sobrequierensubir);
    \path [line] (siYallego) -| node [text width=2.5cm,midway,above=-0.9cm,align=center ] {Yes} (meSirve);
    \path [line] (meSirve) -| node [text width=2.5cm,midway,left=-0.8cm,align=center ] {No} (actualizoUsuarias);
    \path [line] (meSirve) -- node [text width=2.5cm,midway,right=-0.7cm,align=center ] {Yes} (subalo);
    \path [line] (subalo) -| (actualizoUsuarias);
    \path [line] (sobrequierensubir) -| node [text width=2.5cm,midway,left=-0.5cm,align=center ] {Next User}   (quepo);
    \path [line] (subete) -| (sobrequierensubir);
    \path [line] (quepo) -- node [text width=2.5cm,midway,right=-0.7cm,align=center ] {Yes} (subete);
    \path [line] (quepo) --  node [text width=2.5cm,midway,above=-0.8cm,align=center ] {No} (end);
    \iffalse
    \path [line] (sobrequierensubir) -- (end);
\fi
  \end{tikzpicture}}
  \caption{Subroutine: Update Users in Station}
  \label{fig:UpdateUsersInStation}
\end{figure}

This model allows us to estimate how many passengers are there at
every single bus -corresponding to individual services for each line-
and waiting at each of the stations throughout the day. Takes into
account the service schedules and occupation of the buses. 

\subsection{TransMilenio}
\subsubsection{Structure and definitions}
The bogotanian mass transit has the following components \cite{DescripcionTM}:
\emph{troncal}, covered by large buses of either two or three bodies,
capacity of 150 and 270 respectively, using mostly dedicated streets,
and the AFC card validation is done in the station; \emph{zonal},
covered buses with capacity of either 40 or 50 passengers, shares
streets with private traffic, has AFC card validation within the bus;
\emph{alimentador}, uses buses with capacity to move 90 people, covers
short distances in a feeder capacity to the \emph{troncal} system, may
or may not have AFC card validation; \emph{complementario} and
\emph{especial}, with a very few lines, connecting to nearby towns,
with buses of capacity of either 50 or 80 passengers.

The present model is aimed at the \emph{troncal} system, since it is
the largest part of the mass transit system, is already complex enough
as to have a dedicated implementation for itself, and in our view its
optimization has the higher potential to improve the quality of the
service.

%The characteristics of the elements of the network are the following:

\subsubsection{Open Data}
TransMilenio has made public several datasets: AFC card data sets,
that are updated daily; names and locations of the stations; routes
and lines of the system. The data set include the following
information:

\begin{itemize}
\item User card validation: which includes the localization, time, and
  anonymized card ID \cite{ValidacionesTullave}. In some cases users
  are able to use the feeder system without validating. Users do not
  validate egress of the system. There are single usage cards sold at
  the stations as well as multiple usage cards, in which several trips
  are pre-paid. These files are updated daily.
\item Planned bus departure schedules for weekdays and weekends
  \cite{HorariosTM}. This data set not reflect daily service incidents.
\end{itemize}

The TransMilenio data uses the word \emph{line} to denote a list of
stations that can be traveled in both directions; each of which
correspond to a \emph{route}. 

\begin{figure}[htb]
  \includegraphics[width= \columnwidth]{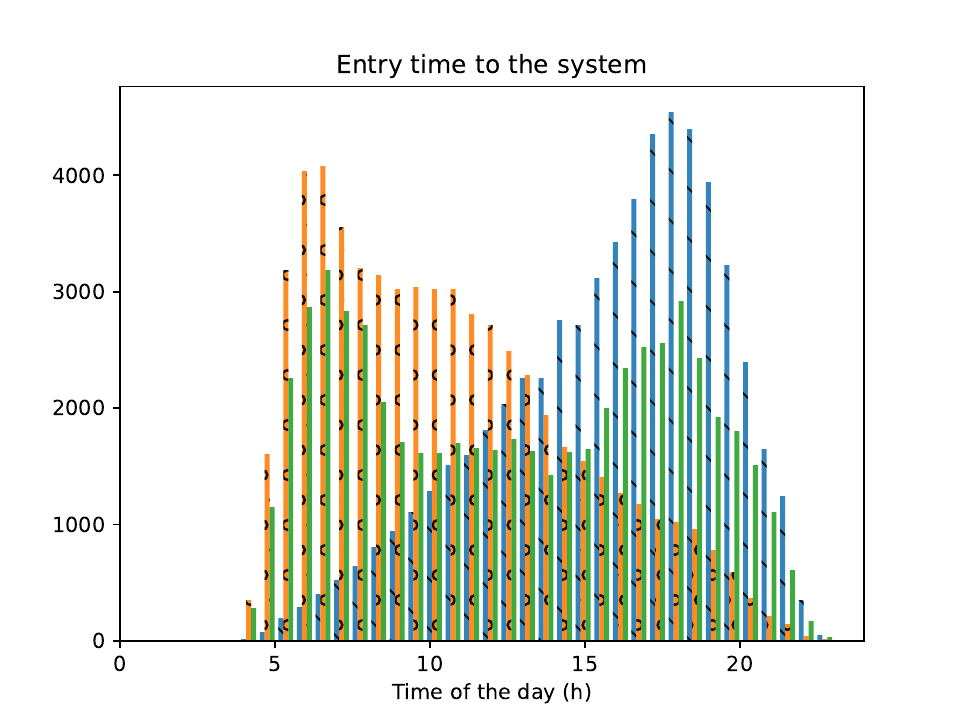}
  \caption{Validation histogram for May 17 of 2020. In yellow (first
    column, hatched) are the first validation of those with more than
    one validation. In blue (third column, circles) the second
    validation of those with more than one validation. In green
    (middle column) the single validations. }
    \label{fig:toe}
\end{figure}

We now discuss the destination inference problem. Remember that for
this situation the BRT systems more closely resemble a metro/subway
rather than a bus system in the sense that we do know which is the
entry station to the system. To tackle the problem we have constructed
histograms of validations from the data set corresponding to May 17 of
2020 Figure \ref{fig:toe}. In this day about a third of the
validations are unique; the card numbers appear only once on the data
set of that day. In Figure \ref{fig:toe} it is represented in the
green histogram. Of the remaining validations, another third
correspond to the first validation of a card that was used more than
one time, and another third to the second validation of those
cards. Notice that the first validation is unimodal, and has its peak is
close to 6 am; likewise the distribution of the second validation has
a peak close to 17:00. Those characteristics are an indication that
those users are following a schedule of 8 am to 5 pm; be it laboral
or educational.

To infer the destination of those trips in which there are two
validations we use follow the procedure described in
\cite{reddy2009entry}. For the first trip the origin corresponds to
the location of the first validation, and the destination to the
location of the second validation; this is coherent with the
distribution of trips shown in the histograms of Figure
\ref{fig:toe}. For the second trip those are reversed, the entry point
to the system is the station of the second validation of the day and
the exit point of the system is the station of the first validation of
the day. This is called the symmetry assumption by Navick et al. 
\cite{doi:10.3141/1799-14}.

\begin{figure}[htb]
  \includegraphics[width= \columnwidth]{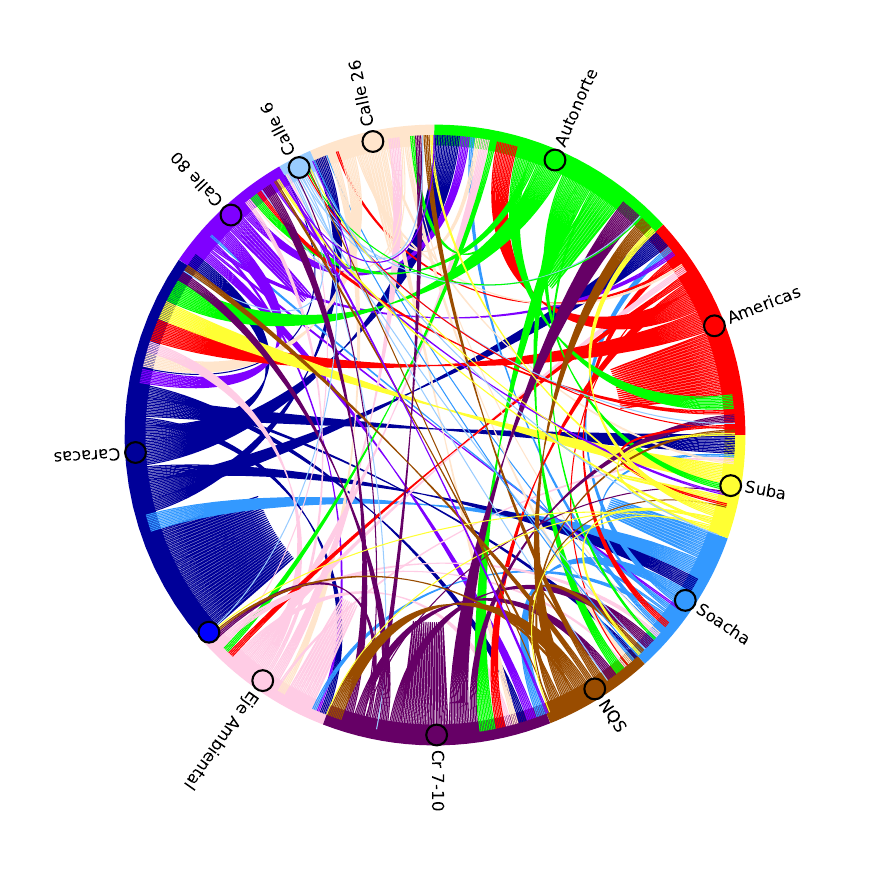}
  \caption{Chord Diagram of the trips between different zones of
    TransMilenio for May 17 of 2020.  Lines join sources with
    destinations, their color corresponds to the source. }
    \label{fig:chord}
\end{figure}

For those cards that have a single validation, we assign the
destination point of a trip with a weighted random function. The
weights of a particular end point correspond to the fraction of the
trips that were inferred from the cards that had more than one
validation. Figure \ref{fig:chord} depicts the chord diagram for the
trips of the day between the different zones of TransMilenio.  The
line joins source zones with destination zones, and its color
corresponds to the source. It can be seen that the \emph{Caracas} zone
is the backbone of the system, having the larger fraction of trips,
and being the endpoint of trips from every other zone. Zones
\emph{Autonorte,Américas,NQS,Cr 7-10, Eje Ambiental, Calle 80, Calle
  26} share a similar behavior: have trips that start and end on the
same zone and are relatively connected by trips with other
zones. \emph{NQS} for its part has a negligible fraction of trips
starting and ending on the same zone. The smallest ammount of trips
correspond to those of the zone \emph{Calle 6}.

\begin{figure}[htb]
 %  https://www.sitp.gov.co/publicaciones/40236/mapas-transmilenio/info/sitp/media/img75080.jpg
  \includegraphics[width= \columnwidth]{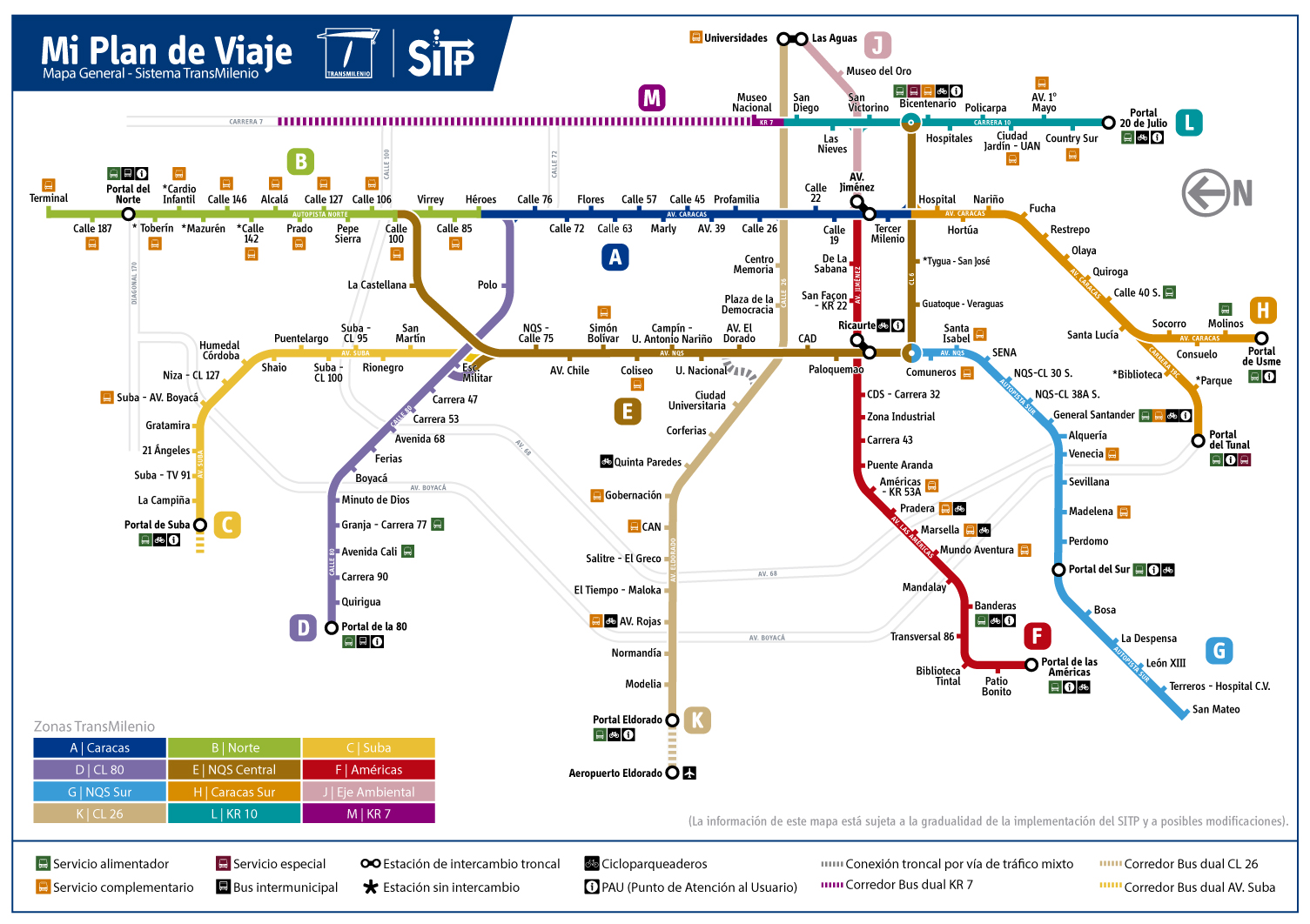}
  \caption{Map of the TransMilenio system. Taken from \url{https://www.sitp.gov.co/publicaciones/40236/mapas-transmilenio/info/sitp/media/img75080.jpg} on Jan 27 2023.
 }
    \label{fig:tmz}
\end{figure}

As a caveat take in mind that there are users who skip the payment by
entering the system through unautorized locations. Therefore, bus
occupation in the system will in general be larger than what we can
predict with our model. Since we do not have reliable data to take
this issue into account we are not going to model it.

\section{Results and Discussion}

Now we will showcase the application to the model to the particular
case of TransMilenio. Fistly we will compare the results of the model
for two particular of the Covid-19 pandemic. The lock-down of the city
provides a way to test whether the system does capture how a
difference in public transport usage translates into different
aggregate conditions for the system. Secondly we analize the
occupation of the system and test different ways in which strategies
of passengers can evolve through a repeated use of the system.

\subsection{Comparison of two different days within the Covid-19 pandemic}
\begin{figure}[htbp]
    \includegraphics[width= \columnwidth]{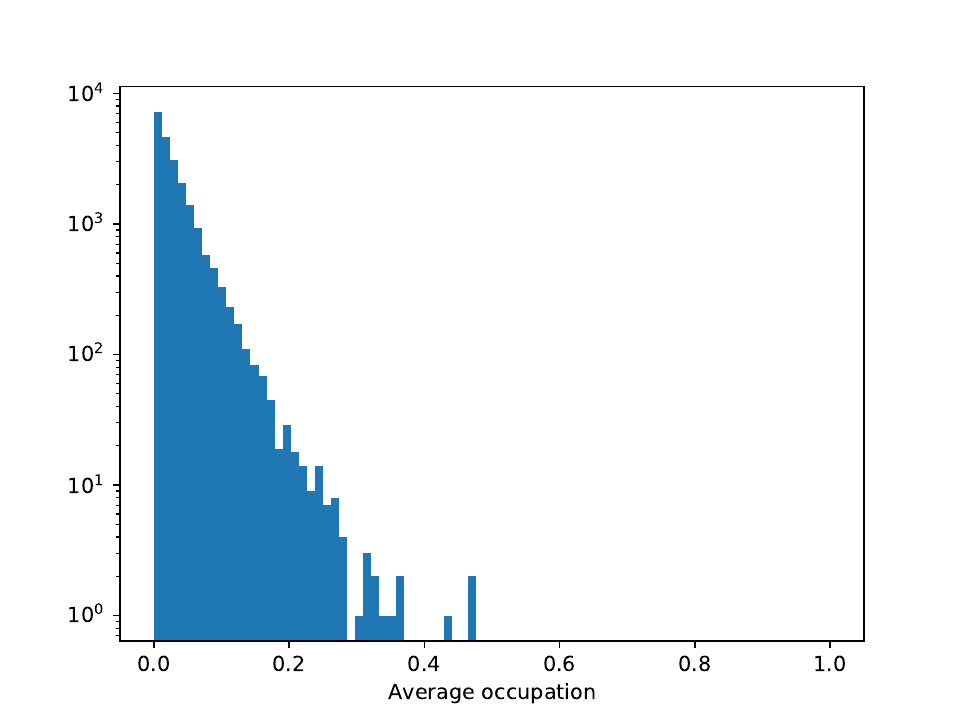}
%  \resizebox{\columnwidth}{!}{}
  \caption{Average occupation of buses on the 17 of May 2020}
  \label{fig:histoccupation20200517}
\end{figure}

Since our simulation follows every passenger throughout the system,
and we run every single instance of a route (what we call a bus), we
are able to find calculate the occupation of every bus. As the
occupation only changes at the stations we have a discrete set of
times at which to calculate the occupation of each bus. Furthermore,
for each bus we can calculate both the average and the maximum
occupation during its transit through a route. Figure
\ref{fig:histoccupation20200517} shows the histogram of average bus
occupation for May 17, 2020. At this point in the COVID-19 pandemic the
usage of public transport in Colombia was down 78\% from pre-pandemic
levels \cite{GoogleMobility,aktay2020google}; which in terms of the
occupation of the buses implied that very few buses had an average
occupation above 40\% and the distribution of bus occupations is peaked
around 0\%. 

\begin{figure}[htbp]
    \includegraphics[width=\columnwidth]{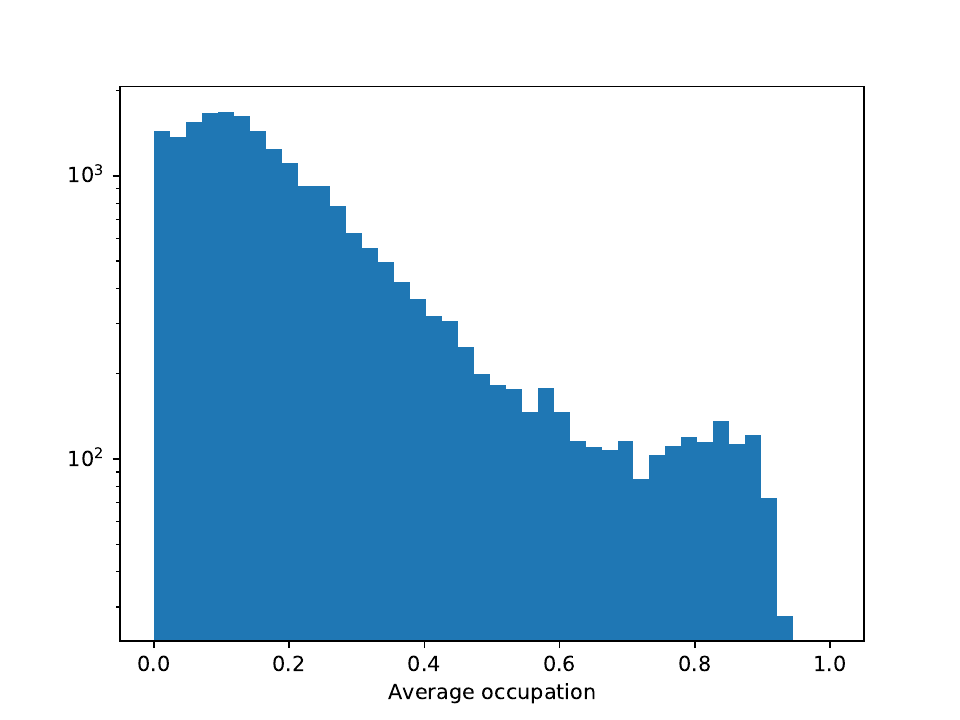}
%  \includegraphics[width=\columnwidth]{graficasOtrasFechas/histoOcupa20201201.png}
%  \resizebox{\columnwidth}{!}{}
  \caption{Average occupation of buses on the 1 of December 2020}
  \label{fig:histoccupation20201201}
\end{figure}

Compare this with the occupation for December 1, 2020, in Figure
\ref{fig:histoccupation20201201}. At this point there was a 15\%
decrece in public transport usage from pre-pandemic levels; thus it
was higher than on the previously studied case. Accordingly, we see
that bus occupations are qualitatively different. Firstly, there are many buses having an average occupation above 60\%, about
1200.

\begin{figure}[htbp]
    \center
  \includegraphics[width= \columnwidth]{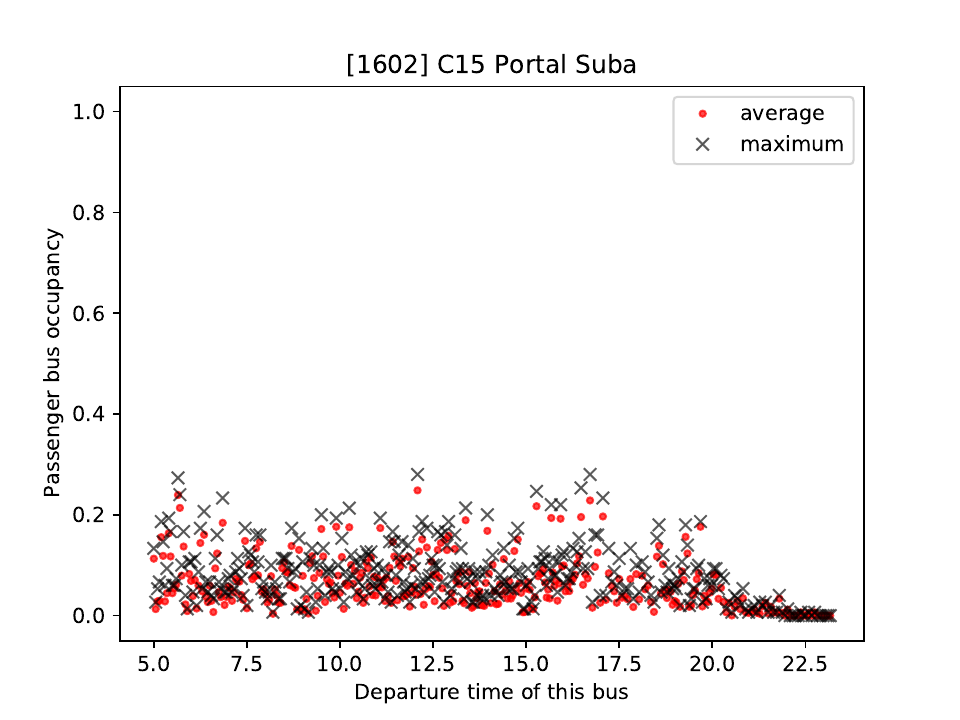}
%  \includegraphics[width= \columnwidth]{data/graficasPP_20200517/001_1602_C15_Portal_Suba.png}
%  \resizebox{\columnwidth}{!}{}
  \caption{Average (circles) and maximum (crosses) occupation of every bus on the routes of the 17 of May 2020 for the line C15 towards Portal Suba.}
  \label{fig:data/graficasPP_20200517/001_1602_C15_Portal_Suba.png}
\end{figure}

\begin{figure}[htbp]
    \center
  \includegraphics[width= \columnwidth]{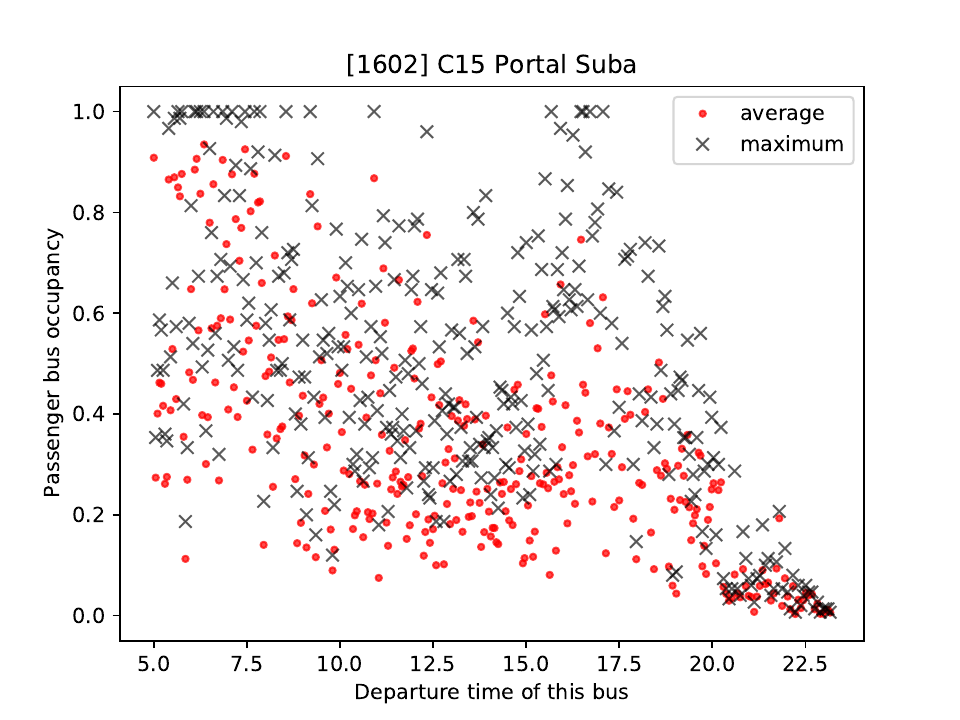}
    \caption{Average (circles) and maximum (crosses) occupation of every bus on the routes of the 1 of December 2020 for the line C15 towards Portal Suba.}
  \label{fig:graficasOtrasFechas/001_1602_C15_Portal_Suba_2020_12_01.png}
\end{figure}

A more detailed description of the bus occupation can be obtained by
taking a particular line. For instance, Figure
\ref{fig:data/graficasPP_20200517/001_1602_C15_Portal_Suba.png}
displays both the average occupation and the maximum occupation of
every bus in the line C15 towards Portal Suba on the 17 of May 2020,
sorted by their departure time. On this date the bus occupancy had a
low variability throughout the day, with a range between 0 and 0.3;
and the average occupancy was close to the maximum occupancy for each
particular bus. Compare with Figure
\ref{fig:graficasOtrasFechas/001_1602_C15_Portal_Suba_2020_12_01.png}
for December 1, 2020. In this case there is in general a larger spread
of both average and maximum occupancy, with a range betwen 0 and 1;
there is a larger gap between any individual bus average and maximum
occupancy; and there is a drop of both quantities for buses starting
their route at 17:30. All those characteristics indicate that
there was an increased use of the line on December 2020 compared to May
2020. This is consistant with the known dynamics of public transport
usage throughout the first year of the Covid Pandemic.

\begin{figure}[htbp]
    \center
  \includegraphics[width= \columnwidth]{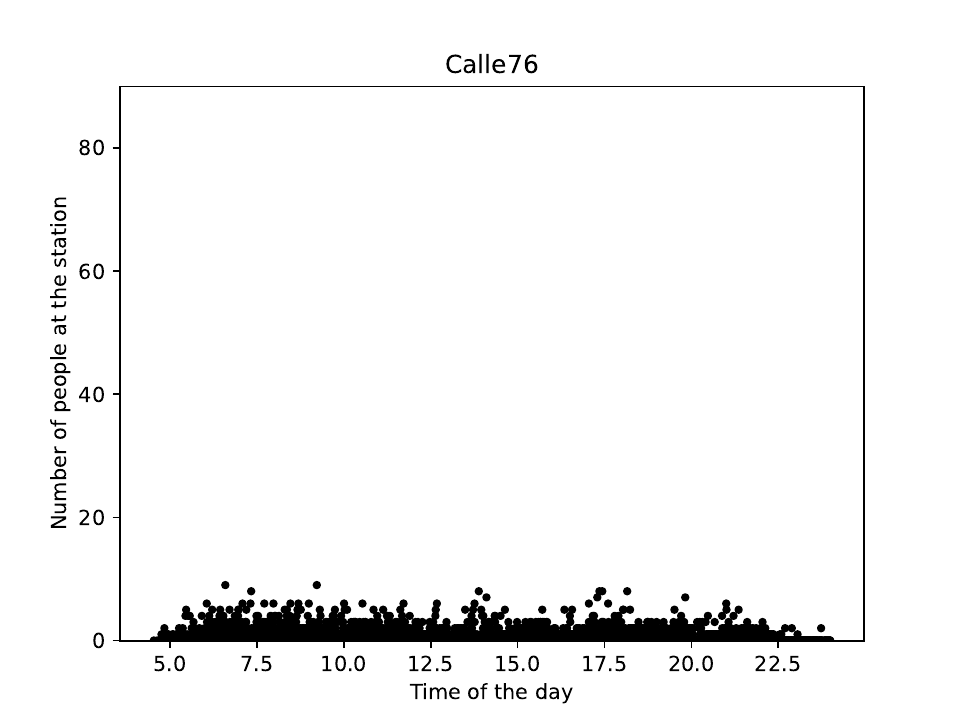}
%/home/gavox/gdrive/03_Investigacion/54_CompartimentosPandemicos/05_codigo_2020_05_17/data/postEstaciones/est9123_Calle76.txt
\caption{Number of people at the station \emph{Calle 76} on May 17, 2020, as a function of time.}
  \label{fig:OcupacionCalle76_17Mayo2020}
\end{figure}

\begin{figure}[htbp]
    \center
  \includegraphics[width= \columnwidth]{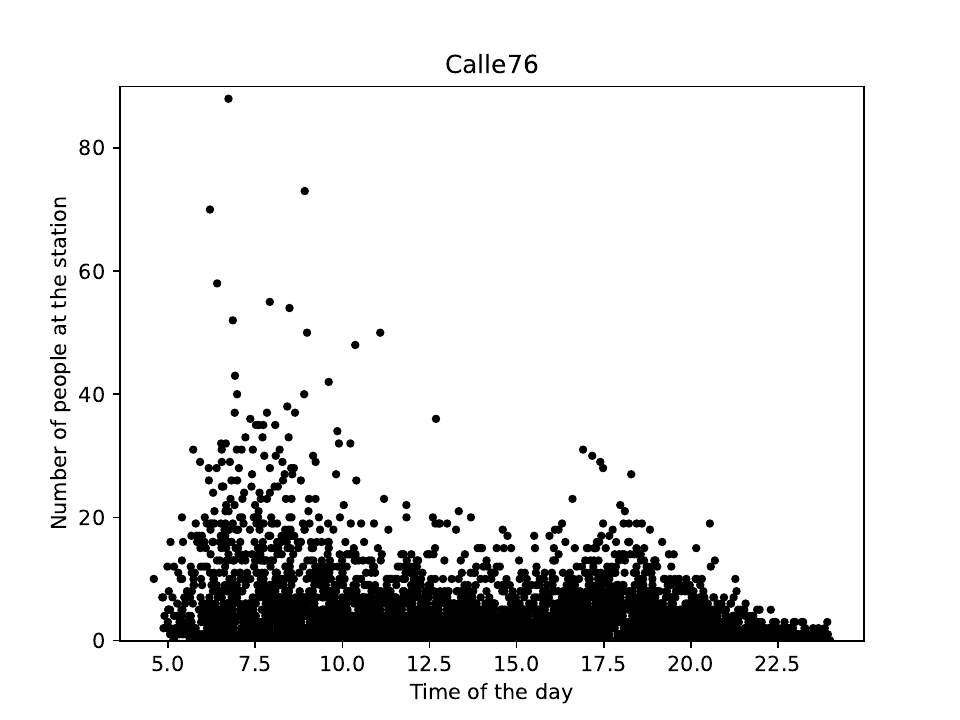}
%/home/gavox/gdrive/03_Investigacion/54_CompartimentosPandemicos/04_codigo_2020_12_01/codigo/data/postEstaciones/est9123_Calle76.txt
\caption{Number of people at the station \emph{Calle 76} on December 1, 2020, as a function of time.}
  \label{fig:OcupacionCalle76_1Diciembre2020}
\end{figure}

We also calculate the number of persons that have arrived to a station
and are waiting for the route. Take for instance the station
\emph{Calle 76}. In Figure \ref{fig:OcupacionCalle76_17Mayo2020} this
quantity takes values smaller than 10 throughout the day of May 17,
2020. Compare with Figure \ref{fig:OcupacionCalle76_1Diciembre2020},
representing December 1, 2020; here the number of people in the station
increases drastically, taking values close to 80, and peaking both at
7:30 and at 15:30.

\section{Conclusions and Closing Remarks}

The situation in which Bogotá transit authorities find themselves with
regard to the ridership of its main public transport system is
certainly not unique; the automatic fare collection (AFC) systems
allow collecting massive amounts of data that nonetheless in its raw
form does not convey the information that can be used to improve the
transport systems.

We have presented a model that takes into account the structure of the
networks, the capacity of the buses, and the AFC data in order to
calculate individual user trajectories and aggregate them; thus
allowing us to make conclusions about the system. Our model can be
applied to improve the system for users, find the points of failure,
and in general understand and model the ridership of the system.

As an application of the model, we have shown how the distribution of
average bus occupation on a particular day of the Covid-19 sanitary
crisis, 17 of May 2020, was remarkably different to a day in which the
mobility restrictions were lifted, 1 of December 2020. This can be
seen as a validation of the model, and an example of how it can be
used to understand the usage of the system and to improve the public
transport experience.

There are several paths that can be taken in order to improve the
present model. Our model implements a simple approach to simulate the
traffic of buses, by having a distribution of speeds. A better model
for the traffic of BRT buses would take into account the traffic
induced by themselves in their restricted lanes, as well as congestion
resulting in queues in accessing the stations. We plan to include these
in future iterations of the model.

Although we are confident that we have a solid basis to
infer the destination of the trips, a possible way to reduce the
uncertainty would be to have direct information of the exit station. A
Big Data approach to this problem will be to use the location of
individual cellphones of users out of cell network provider
information; this presents a more precise location of each passenger,
but may include a sampling bias if there are technological differences
among the usage of cellphone and coverage of service related to the
income of passengers.

\section*{Contributions}
A.A., J.G., and G.V. worked on the conceptual structure of the
model. J.G. wrote the network model for TM. A.A. and G.V. wrote the
computer code for the users. G.V. and J.G. cleaned the data, and wrote
the article. G.V. ran the simulations and post-processed the data.

\section*{Competing Interests and Funding}

The present work lies within the research project ID E5 2023 8,
\emph{Escuela Superior de Administración Pública.} Besides that, it
did not receive any specific grant from funding agencies in the
public, commercial, or not-for-profit sectors. 

\bibliographystyle{alpha}
\bibliography{bibcp}

%\end{multicols}

\end{document}